\documentclass[journal]{IEEEtran}

\usepackage[dvips]{graphicx}
\graphicspath{{./}}
\DeclareGraphicsExtensions{.eps}
\usepackage[cmex10]{amsmath}
\begin{document}

\title{Two Step SOVA-based Decoding Algorithm for Tailbiting Codes}

\author{Jorge~Ort\' in,~\IEEEmembership{Student~Member,~IEEE,}
       Paloma~Garc\' ia,~Fernando~Guti\' errez~and~Antonio~Valdovinos%
\thanks{The authors are with the Arag\' on Institute for Engineering Research (I3A), University of Zaragoza, Zaragoza E-50018, Spain.

This work has been financed by the Spanish Government (Project
TEC2008-06684-C03-02/TEC from MCI and FEDER),
Gobierno de Arag\'on (Project PI003/08 and WALQA Technology Park) and the
European IST Project EUWB.}}

\maketitle

\begin{abstract}
In this work we propose a novel decoding algorithm for tailbiting convolutional codes and evaluate its performance over different channels. The proposed method consists on a fixed two-step Viterbi decoding of the received data. In the first step, an estimation of the most likely state is performed based on a SOVA decoding. The second step consists of a conventional Viterbi decoding that employs the state estimated in the previous step as the initial and final states of the trellis. Simulations results show a performance close to that of maximum-likelihood decoding.
\end{abstract}


\section{Introduction}
\IEEEPARstart{T}{ailbiting} convolutional codes are those whose initial and final states are constrained to be the same, forcing every codeword to start and finish at identical states. Unlike Zero Tailing codes, where each data block is followed by a zero tail which sets a known final state, tailbiting codes preset the encoder memory with the last bits of the block, being these bits unknown to the decoder. While this method improves the efficiency of the encoding, it also increases the complexity of the decoding process \cite{Tail:Cox}. 

Maximum Likelihood (ML) decoding of tailbiting codes requires to perform $2^{M}$ separate Viterbi decodings, each of them starting and finishing with one of the $2^{M}$ possible initial and final states, to select the decoded path with the best metric. Owing to the large burden of ML decoding, alternative decoding algorithms have been proposed in the literature, many of which are based on the Circular Viterbi algorithm (CVA) proposed in \cite{Tail:Cox}. This decoding method exploits the circularity of tailbiting codes concatenating the same received block several times and extending the Viterbi algorithm around them until a stopping rule is fulfilled. Several modifications of this scheme have been proposed to improve its performance in terms of both BER and convergence \cite{Tail:Shao}. Recently, a new algorithm has been proposed in \cite{Tail:Pai} which reduces significantly the computational complexity of Maximum Likelihood decoding, thus allowing its implementation for real systems. 

Although many of the previous algorithms present terminating conditions to avoid situations of non converge of the decoding process, the number of operations required to decode a block is generally variable and a function of the Signal to Noise Ratio (SNR). The use of tailbiting codes in mobile systems with rapid variations in their channel conditions causes the presence of received blocks which can be affected by severe degradation. This fact can worsen the convergence of the CVA-based algorithms and increase the variability of the processing time required to decode each block. In those cases, the development of a decoding algorithm whose decoding time is fixed for all the SNR and channel conditions and whose results are also near to those of the ML decoding would simplify the design and implementation of the receiver improving its performance.

In this work, we propose a two-step decoding algorithm for tailbiting convolutional codes. Assuming that all the possible states are equally probable to be the initial state, the first step of the algorithm searches a state in the trellis whose likelihood is maximum. This search is done with a modified version of the Soft-Output Viterbi Algorithm (SOVA) \cite{Decoding:Hagenauer}. Once this state is chosen, the circular features of tailbiting codes are employed and a second Viterbi decoding is performed starting and finishing at the selected state in the previous step.

The performance of the proposed algorithm, which we will call two-step Viterbi algorithm (TSVA), is evaluated over the additive white Gaussian noise (AWGN) channel and over a wide sense stationary uncorrelated scattering (WSSUS) channel assuming that it is employed in an OFDM system. This kind of scenario is typical of cutting edge mobile systems which use tailbiting codes, such as WiMAX \cite{general:802.16} or LTE \cite{encoding:LTE}. In that sense, this election allows evaluating the performance of the TSVA in real operation conditions.  

This work is organized as follows. Section II gives a detailed explanation of the operation of the TSVA, specifying its different steps. In section III, simulation results are presented and the computational load of the TSVA is estimated. Finally, the conclusions are summarized in section IV.

\section{Algorithm Description}
As stated in the previous section, CVA decoding ties around the final of the block with its beginning until a stable decoded path is achieved. Nevertheless, this strategy can lead to the appearance of non converging blocks and \textit{pseudocodewords}, that is, tailbiting paths whose recurrence is higher than one trellis. Unlike the CVA, the TSVA tries to choose a proper initial and final state for the trellis instead of looking for a stable decoded path. In that sense, the TSVA prevents the presence of \textit{pseudocodewords}, but it can present additional errors caused by a bad selection of the initial and final states. The principal aim of the algorithm is to decrease as much as possible this error. 

In order to obtain a proper initial and final state, we need to select a state of the trellis whose likelihood is maximum. For this purpose we use a modified version of the SOVA \cite{Decoding:Hagenauer} to obtain the likelihood of the states in a specific path. Suppose we have received an encoded block with soft information from the demodulator. Since there is no \textit{a priori} information regarding the initial conditions of the trellis, we start with all the states having a metric equal to zero. As the calculation of the trellis evolves, we also record for each state $s_k$ at a given time instant $i$ the soft metric $\Delta(s_k^i)$, defined as:
\begin{equation}
\Delta(s_k^i) = \max(\Gamma(s_{p}^{i-1},s_{k}^{i}))-\min(\Gamma(s_{q}^{i-1},s_{k}^i))
\end{equation}
where $\Gamma(s_{p}^{i-1},s_{k}^{i})$ is the length (accumulated metric) of the path which merges in state $s_k$ at the time $i$ through state $s_p$, i. e., its last branch corresponds to a transition from state $s_p$ to state $s_k$. The survivor path is the one corresponding to $\min(\Gamma(s_{p}^{i-1},s_{k}^{i}))$ as defined in the classical Viterbi algorithm. This soft metric is closely related to the likelihood of the survivor path, since high values of $\Delta(s_k^i)$ corresponds to low probability of selecting the wrong survivor path.

When finished the trellis calculation, the survivor path with minimum length is chosen to traceback, even if this path is not tailbiting. The likelihood of each state $L(s_k^i)$ in this path is obtained by selecting the minimum of all the recorded $\Delta(s_l^j)$ in the path, with $j \geq i$, corresponding to discarded paths not passing across that state. 

\begin{figure}[!t]
\centering
\includegraphics[width=3.8in, clip=true]{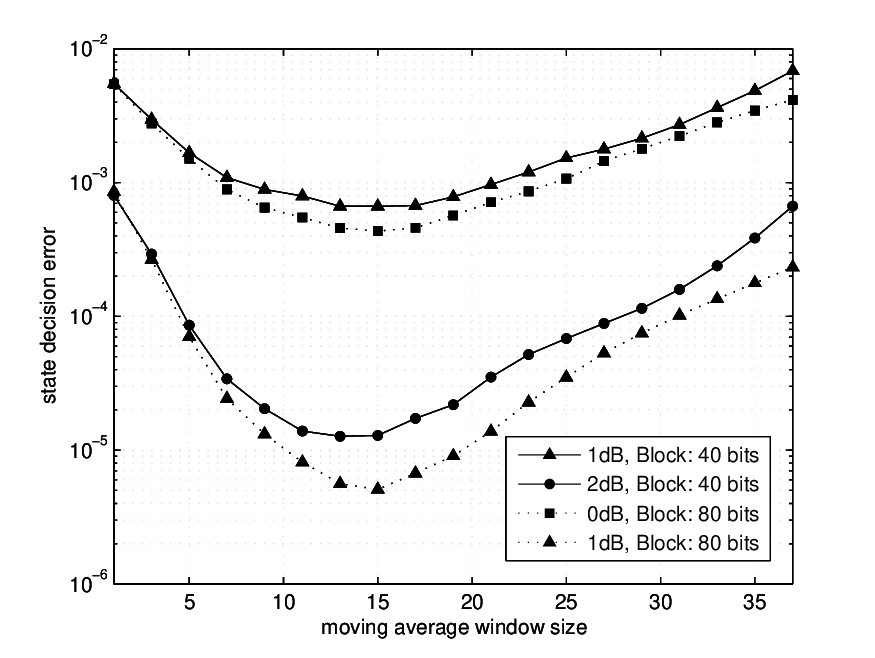}
\caption{State decision error of the (3,1,6) tailbiting code with generators (171,133,165) as a function of the window size with different SNRs and block sizes over an AWGN channel.}
\label{fig:window_size}
\end{figure}

Once the likelihoods of each state in the minimum length path have been estimated, the most likely state could be selected by only searching the state $s_k^i$ at which $L(s_k^i)$ is maximum. Nevertheless, this method employed to obtain the likelihoods can lead to the presence of consecutive states in the path with disparate likelihoods due to estimation errors. In that sense, considering the estimated likelihoods of the surrounding states can improve the reliability of the choice. In order to introduce these likelihoods, we propose the use of a moving average applied to the sequence of estimated state likelihoods along the path:
\begin{equation}
\hat{L}(s_k^i) = \sum_{m=0}^{M-1}{L(s_l^{i+m})}
\end{equation}
where $\hat{L}(s^i_k)$ is the averaged estimated likelihood of each state in the path and $M$ is the window size, which is set heuristically to minimize the error in the selection of the initial state for the second Viterbi decoding. As an example, fig. \ref{fig:window_size} shows the error in the choice of the initial state for the second Viterbi decoding as a function of the window size with different SNRs and block sizes over an AWGN channel for the (3,1,6) tailbiting code with generators (171,133,165). This error is defined as the ratio of the number of incorrect decisions for the selected state (i.e. the selected state does not correspond to a state of the transmitted sequence) to the total number of transmitted blocks. The results show the presence of an optimum window size which is independent of the SNR and the encoded block size. Similar results can be found for different encoders.

Once an initial state has been chosen, the circular properties of the tailbiting codes allow arranging the received sequence and starting a second Viterbi decoding considering the selected state as the beginning and end of the trellis. The rearranged output of this second Viterbi decoding is the final decoded data. In those cases where a very short block is employed, it is possible to concatenate several times the same block and perform the search in the resulting trellis. The following steps summarize the TSVA:

\begin{enumerate}
    \item \textit{Start the decoding from all the initial states in the trellis with the metric state set to zero.}
    \item \textit{Process the trellis to the final states, storing in each update the term $\Delta(s_k^i)$ for all the states.}
    \item \textit{Choose the final state with minimum accumulate metric and traceback, recording the likelihood of the states $L(s_k^i)$ in the path. This path may not be tailbiting.}
    \item \textit{Apply the moving average to the sequence of likelihoods of the selected path, obtaining the sequence $\hat{L}(s_k^i)$.}
    \item \textit{Search the position $i$ at which $\hat{L}(s_k^i)$ is maximum and store the state at that position of the path.}
    \item \textit{Rearrange the sequence and start a second Viterbi decoding considering the state stored in the previous step as the initial and final state of the trellis.}
\end{enumerate} 

\begin{figure}[!t]
\centering
\includegraphics[width=3.8in, clip=true]{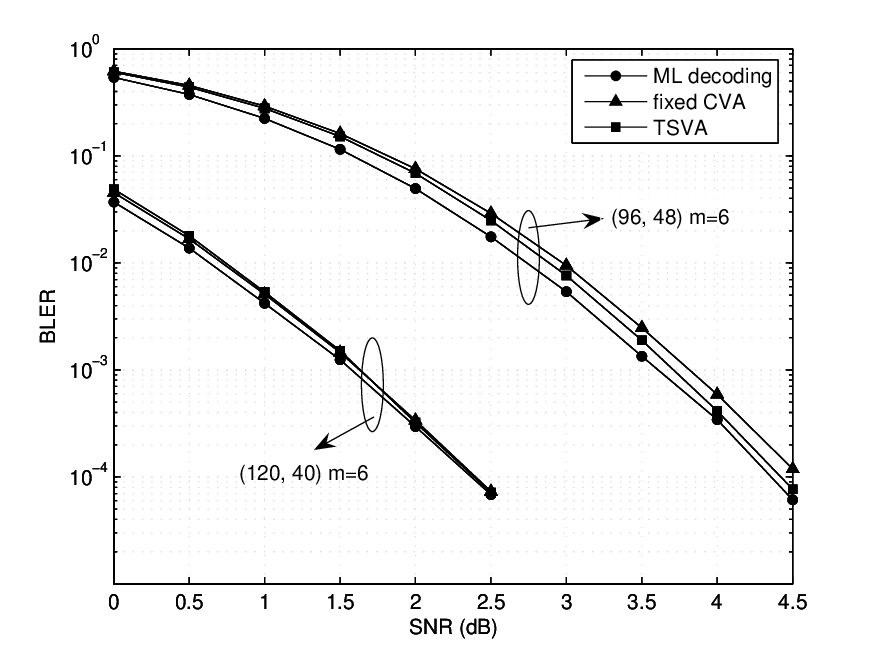}
\caption{Block Error Rate (BLER) of the (96,48) and the (120,40) tailbiting codes over the AWGN channel decoded by various algorithms with soft decision decoding.}
\label{fig:BLER_AWGN}
\end{figure}

\section{Simulation Results}
In this section, we show the performance of the TSVA algorithm in an OFDM system over the AWGN and the ITU Vehicular A channels. This choice aims at simulating the performance of the proposed algorithm considering the technology and channel conditions which will be found in new mobile technologies with a tailbiting coding scheme. 

The simulated OFDM signal has a bandwidth of 5 MHz, the carrier frequency is 3.5 GHz and the symbol period is 102 $\mu$s. The number of subcarriers in a symbol is 512, being the subcarrier space 10.9 KHz. The mobile speed is 120 km/h, leading to a Doppler frequency of 389 Hz. The considered codes are the (96,48) and (120,40) tailibiting codes with generators (171,133) and (171,133,165) respectively. These coding schemes are proposed to be used in the 802.16e \cite{general:802.16} and LTE \cite{encoding:LTE} standards. The encoded data are interleaved and mapped to QPSK symbols. It has been assumed an ideal channel estimation and soft-decoding at the receiver. The results have been obtained using Montecarlo method, ensuring at least 100 reported errors for every simulation result.

\begin{figure}[!t]
\centering
\includegraphics[width=3.8in, clip=true]{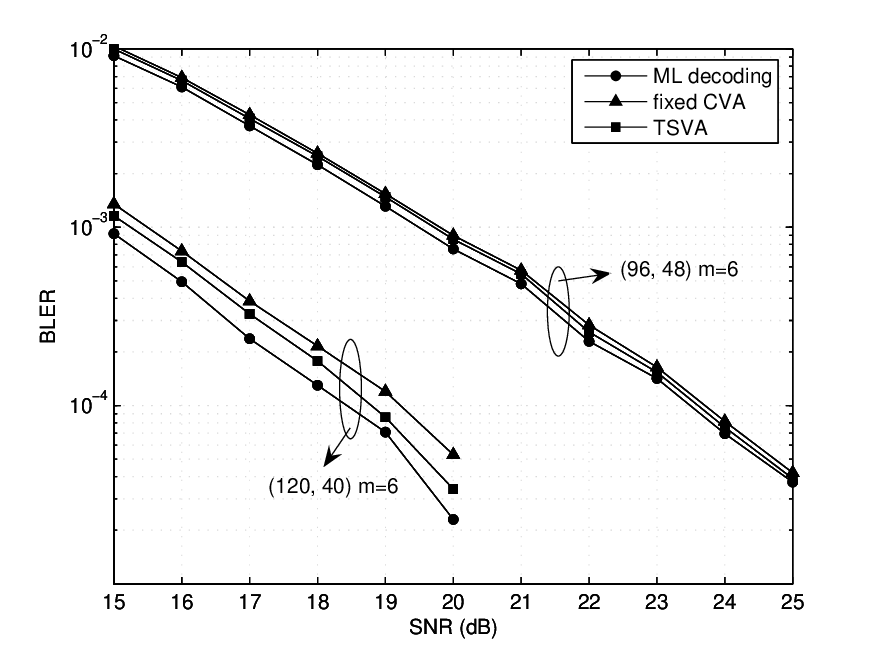}
\caption{Block Error Rate (BLER) of the (96,48) and the (120,40) tailbiting codes over the ITU Vehicular A channel decoded by various algorithms with soft decision decoding.}
\label{fig:BLER_WSSUS}
\end{figure}

In Fig. \ref{fig:BLER_AWGN}, the Block Error Rate (BLER) of the TSVA in AWGN conditions is compared with those obtained by the ML decoding and the CVA when its length is constrained to two blocks. This fixed CVA is equivalent in terms of Viterbi updates to the TSVA as it will be demonstrated in the next paragraphs. The results attain a near-optimum performance, with a slight penalty (below 0.2dB) with respect to ML decoding and improving the results obtained with the fixed CVA. Fig \ref{fig:BLER_WSSUS} shows similar results when the BLER is obtained over an WSSUS channel. In this case, the improvement caused by the use of the TSVA instead of the fixed CVA is higher since the TSVA can avoid zones of the trellis with error bursts and low likelihoods for the selection of the initial state.

In order to estimate the computational load of the TSVA, it is necessary to calculate the total number of Viterbi updates required for the decoding of a block and the extra operations in each Viterbi update for the estimation of the initial state. These additional operations only affect the first phase of the algorithm, since likelihood estimations are not needed in the second pass. 

Typically, each Viterbi update consists of an addition and a comparison for all of the possible states in the trellis. The required operations for the state estimation is essentially an additional subtraction for each state when computing the Viterbi update. The comparison can be easily done considering the sign of the subtraction and therefore it can be suppressed. In that sense, the computational load of each Viterbi update in the TSVA remains the same. Furthermore, the TSVA requires an additional traceback and the averaging of the state likelihood in the selected path, but these additional operations are negligible in comparison with the computation of the Viterbi updates in the trellis. The total number of Viterbi updates is $(N+1)N_{Block}$, being $N_{Block}$ the size of the encoded block and $N$ the number of times that it is concatenated in the decoding process. Unlike other proposed algorithms, this result is constant for all the SNRs and completely determinist.

\section{Conclusion}
A new tailbiting decoding algorithm is proposed and its performance is evaluated over different channels. The algorithm is composed of two differentiated steps. In the first step a estimation of the likelihood of the states in the most likely path is done. Based on this estimation and thanks to the circular properties of the tailbiting codes, an initial state is chosen in order to perform a new Viterbi decoding in the second step of the algorithm. In this case, the trellis is forced to start and finish in the state selected previously. Simulation results show that the proposed algorithm achieves near-optimum performance in terms of BLER with a fixed and relatively low computational load for the whole range of possible SNR.

\ifCLASSOPTIONcaptionsoff
  \newpage
\fi

\bibliographystyle{IEEEtran}
\bibliography{./IEEEabrv,./bibliografia}
\end{document}